# First results of the SOAP project.
# Open access publishing in 2010

September 14th, 2010

*Suenje Dallmeier-Tiessen [a], Robert Darby [b], Bettina Goerner [c], Jenni Hyppoelae [a], Peter Igo-Kemenes [a #], Deborah Kahn [d], Simon Lambert [b], Anja Lengenfelder [e], Chris Leonard [d], Salvatore Mele [a,*], Panayiota Polydoratou [e], David Ross [f], Sergio Ruiz-Perez [a], Ralf Schimmer [e], Mark Swaisland [g] and Wim van der Stelt [h]*

[a] CERN, CH1211, Geneva 23, Switzerland
[b] STFC Rutherford Appleton Laboratory, Harwell Science and Innovation Campus, Didcot OX11 0QX, United Kingdom
[c] Springer-Verlag, GmbH, Tiergartenstrasse 17, 69121 Heidelberg, Germany
[d] BioMed Central, 236 Gray's Inn Road, London WC1X 8HL, United Kingdom
[e] Max Planck Digital Library, Amalienstr. 33, 80799 Munich, Germany
[f] SAGE, 1 Oliver's Yard, 55 City Road, London, EC1Y 1SP, United Kingdom
[g] STFC Daresbury Laboratory, Warrington, Cheshire WA4 4AD, United Kingdom
[h] Springer Science+Business Media B.V., Van Godewijckstraat 30, 3311 GX Dordrecht, Netherlands
[#] Also at Gjøvik University College, Po.box 191 Teknologivn. 22, 2802 Gjøvik, Norway
[*] Corresponding author: Salvatore.Mele@cern.ch

**Abstract**

The SOAP (Study of Open Access Publishing) project has compiled data on the present *offer* for open access publishing in online peer-reviewed journals. Starting from the Directory of Open Access Journals, several sources of data are considered, including inspection of journal web site and direct inquiries within the publishing industry. Several results are derived and discussed, together with their correlations: the number of open access journals and articles; their subject area; the starting date of open access journals; the size and business models of open access publishers; the licensing models; the presence of an impact factor; the uptake of hybrid open access.

## 1. Introduction

The SOAP (Study of Open Access Publishing) project[1] describes and compares the *offer* and *demand* for open access publishing in peer-reviewed journals. The project has three phases. It first describes the *offer* in open access publishing solutions, which is discussed in this document. The *demand* is then assessed by a large-scale survey of scientists across disciplines and around the world. The *offer* and the *demand* are compared in the final phase of the project. This article describes the first results of the project and presents the most detailed quantitative description to date of the landscape of existing open access journals and publishers. It captures their similarities and differences, volume of publication and business models, evolution with time and subject area. It starts from information available in the Directory of Open Access Journals (DOAJ) [1], complemented by data from other sources, including an inspection of web sites of publishers and journals. At the same time, the market penetration of the hybrid open access publishing model is assessed. This model corresponds to the possibility for articles to be published open access in subscription journals, against the payment of a fee or other agreements between publishers and authors. Section 2 presents the methodology used for collecting the data used in the analysis. Section 3 presents results on the key indicators for the classification of Open Access journals. Section 4 summarises the results in what we hope will provide a fact-based and impartial contribution to the open access debate.

---

[1] The project is financed by the European Commission under the Seventh Framework Programme, and runs from March 2009 to February 2011. The project is co-ordinated by CERN, the European Organization for Nuclear Research, and is a partnership of publishers (Springer, Sage, BioMed Central), libraries (the Max Planck Digital Library of the Max Planck Society) and funding agencies (the UK Science and Technology Facilities Council). For further information: http://soap-fp7.eu

## 2. Methodology

### Quantitative analysis of open access journals

Data describing open access journals and their publishers were downloaded from the DOAJ in July 2009 and then enriched by using data present in August 2009 in the Electronic Journals Library (EZB) [2], SCOPUS [3]; the Journal Citation Reports (ISI-JCR) [4] and SCImago [5]. SOAP partners BioMed Central and SAGE (on behalf of Hindawi Publishing Corporation) provided further data. Additional information relevant for this study was manually collected between September 2009 and January 2010 from the websites of relevant journals and publishers. A detailed description of the methodology and the results is given in Reference [6]. A description of the individual fields used in this study, together with the source of data (DOAJ, other databases, or manually collected), is given in the following.

- *Title (DOAJ).* Title of the journal, as recorded in the DOAJ.
- *ISSN (DOAJ).* ISSN of the journal, as recorded in the DOAJ.
- *Homepage (DOAJ, with some modification).* Journal URL, as recorded in the DOAJ. In some cases these were manually corrected.
- *Start date (DOAJ).* Year of publication of the earliest available open access online content. This refers not only to new open access journals, but also to journals that have made earlier content open access.
- *End date (automatically matched or manually collected).* Year when journal ceased as open access publication. The information was extracted from EZB through the journal ISSN. Manual verification was necessary for a large number of cases.
- *Language (DOAJ, with some manual modification).* The language(s) of the journal. For some journals the putative publication in English language was manually verified.
- *Subject discipline (DOAJ, with some manual modification).* There are 17 subject categories listed in the DOAJ. For the purposes of this study, these are aggregated in six major categories: Chemistry, Physics and Technology; Biology and Life Sciences; Medicine and Health Sciences; Social Sciences; Humanities; General Works.
- *Number of Articles per journal and year (automatically matched and manually collected).* This information was extracted from ISI-JCR and SCImago, starting from the ISSN of the journal. If not available from either of these sources the number of articles was collected manually from the journal websites. This extraction was performed for the year 2008 or, in case this was not conclusive enough for the frequency of issues or articles, for the year 2007. SOAP project partners, BioMed Central and SAGE UK (on behalf of Hindawi Publishing Corporation) directly provided the number of articles for the journals they publish.
- *Income sources (manually collected).* An inspection of the web pages of each journal suggested the different possible income sources: article processing charge, membership fee, advertisement, sponsorship, subscription (to the print version of the journal), sales of hard copies, and other sources (such as print based-colour page charge, off-prints and reprints sales, print on demand, income via conference fees, donations, services to authors).

- *Copyright / licensing options (manually collected).* The information about the copyright/licensing options of the journal was collected manually from the journal websites and assigned to one of three categories: author retains copyright; copyright transfer; a Creative Commons license [7].

Data imported from the DOAJ comprised 4,032 unique journals from 2,588 unique publishers. Journals not in English (25%) and those having ceased publication (a further 7%) were removed. The final sample comprises 2,838 journals by 1,809 publishers. The number of articles per journal and per year (using in some cases data from 2007 or 2008, or the average of the two) was counted for 2,711 journals (96%). It sums up to a total of 116,883 articles.

## The uptake of the hybrid model

The hybrid model for open access publishing was introduced in 2004 and spread quickly: a list of publishers offering this option [8] counted 80 publishers in the beginning of 2010. In order to assess the market penetration of this business model, it is necessary to understand how many articles are published in journals with this option, while keeping into account the overall volume of articles published in the same journals. Most publishers offering this option are relatively small operations, with a few journals and articles/year, so contribute marginally to this overall statistics. On the other hand, twelve publishers offering the hybrid option are instead large publishing houses and their study is sufficient to assess the uptake of the model. At the time of the study, mid-2009, these were: the American Chemical Society, American Physical Society, Cambridge University Press, Elsevier (including Cell Press), Nature Publishing Group, Oxford University Press, PNAS, the UK Royal Society, SAGE, Springer, Taylor & Francis and Wiley Blackwell. These publishers were directly approached with inquiries about the number of articles published with the hybrid option and the total number of articles published in journals with and without the hybrid option.

# 3. Results

## Publisher and journal sizes

The distribution of journals per publisher, presented in Table 1, is highly skewed: on one hand, the vast majority of publishers have only one open access journal; on the other, five publishers have more than 50 journals each, altogether representing 19% of journals and 13% of articles/year. Most publishers (~90%) publish less than 100 articles/year and altogether contribute one third of the total number of articles/year. The remaining two thirds of articles/year are published by the remaining 10% of publishers.

| Journals/Publisher | Publishers | | Total journals | | Total articles/year | |
|---|---|---|---|---|---|---|
| 1 | 1,621 | 90 % | 1,621 | 57 % | 63,887 | 55 % |
| 2 to 9 | 171 | 9 % | 491 | 17 % | 25,442 | 22 % |
| 10 to 49 | 12 | 1 % | 190 | 7 % | 12,623 | 11 % |
| ≥ 50 | 5 | <1 % | 536 | 19 % | 14,931 | 13 % |
| Total | 1,809 | | 2,838 | | 116,883 | |

*Table 1: "Size" of publishers by number of journals and articles*

Two vastly different groups of publishers exist: 14 "large publishers"[2] and "other publishers". These "large publishers", listed in Table 2, publish a total of 36,096 articles/year in 616 journals, 30% of the total yearly output. Six of them are commercial, six are non-profit organisations[3].

| Publisher name | Journals | Articles/year | Publisher type |
| --- | --- | --- | --- |
| BioMed Central | 176 | 8,993 | publishing house commercial |
| International Union of Crystallography | 1 | 5,165 | other non-profit |
| Public Library of Science | 7 | 4,368 | publishing house non-profit |
| Asian Network for Scientific Information | 13 | 2,514 | publishing house commercial |
| Hindawi Publishing Corporation | 85 | 2,044 | publishing house commercial |
| Copernicus Publications | 18 | 2,012 | publishing house non-profit |
| Optical Society of America | 1 | 1,961 | learned society non-profit |
| World Academy of Science, Engineering and Technology | 18 | 1,960 | other N/A |
| Bentham Open | 154 | 1,663 | publishing house commercial |
| Medknow Publications | 59 | 1,574 | publishing house commercial |
| Indian Academy of Sciences | 10 | 1,152 | learned society non-profit |
| Oxford University Press | 2 | 1,032 | other non-profit |
| Academic Journals | 10 | 1,001 | publishing house N/A |
| Internet Scientific Publications | 62 | 657 | publishing house commercial |

*Table 2: The 14 "large publishers", ordered by number of articles per year*

## Subject categories

Table 3 presents the distribution of publishers, journals and articles by subject categories. With the exception of general, multidisciplinary, titles, the distribution over disciplines is rather even. Grouped together, the science technology and medicine fields (STM) (cpt, bio and med) comprise two thirds of the journals and more than three quarters of the articles. Social sciences and humanities (SSH) (soc, hum) comprise 32% of journals and 16% of articles. The output of the "large publishers" is almost exclusively in STM: SSH represents only 5% of their journals and less than 1% of their articles.

---

[2] A publisher is defined as "large" if it published either more than 50 journals or more than 1,000 articles in 2007 or 2008.
[3] The precise status of the other two large publishers, Academic Journals and World Academy Science, Engineering & Technology, could not be immediately determined.

| Subject category | Publishers[4] | | Journals | | Articles/year | |
|---|---|---|---|---|---|---|
| cpt – Chemistry, Physics and Technology (includes Mathematics and Astronomy) | 360 | 20% | 549 | 19% | 33,158 | 28% |
| bio – Biology and Life Sciences | 355 | 20% | 533 | 19% | 24,767 | 21% |
| med – Medicine and Health Sciences | 406 | 22% | 806 | 28% | 32,879 | 28% |
| soc – Social Sciences | 533 | 29% | 611 | 22% | 13,506 | 12% |
| hum – Humanities | 258 | 14% | 276 | 10% | 5,030 | 4 % |
| gen – General works | 63 | 3% | 63 | 2% | 7,543 | 6% |
| Total | 1,809 | | 2,838 | | 116,883 | |

*Table 3: Distribution of publishers, journals and articles by subject category.*

## Journal starting dates

Figure 1 presents the starting date of journals and their subject area. A first period of growth corresponds to the onset of electronic publishing, in the mid-1990s. Approximately 200-300 new titles were added annually in recent years, mostly in the medical sciences by large publishers. Many journals in Chemistry, Physics and Technology, mostly from the other publishers have earlier starting dates.

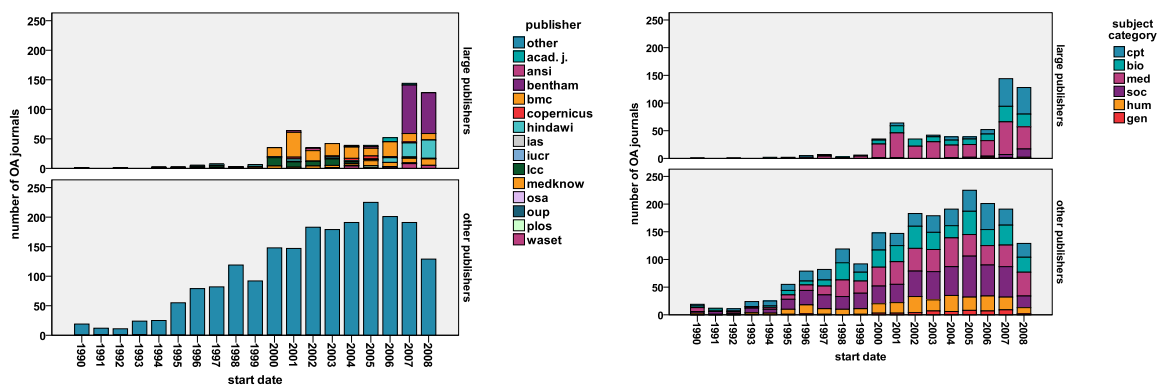

*Figure 1: Journal start date by publisher and subject area.*

## Income sources/business models

Several possible income sources/business models exist for open access journals: article processing charge, membership fee, advertisement, sponsorship, subscription (to the print version of the journal), sales of hard copies, and other sources (such as print based-colour page charge, off-prints and reprints sales, print on demand, income via conference fees, donations, services to authors).

Information on income sources/business model was found for almost all journals of the large publishers but could only be retrieved for about 60% of the journals of other publishers. The results are presented in Figure 2. Article processing charge, membership fees and advertisements are the predominant sources of income for large publishers. For other publishers, sponsorship and print subscriptions are the most important. In addition, there seems to be some correlation between subject domain and income sources for the "other publishers": article processing charges and subscription appears to be favoured for STM titles, while SSH titles seem to favour sponsorship.

---

[4] The same publisher may publish journals in more than one subject areas and publisher figures in this table represent multiple entries.

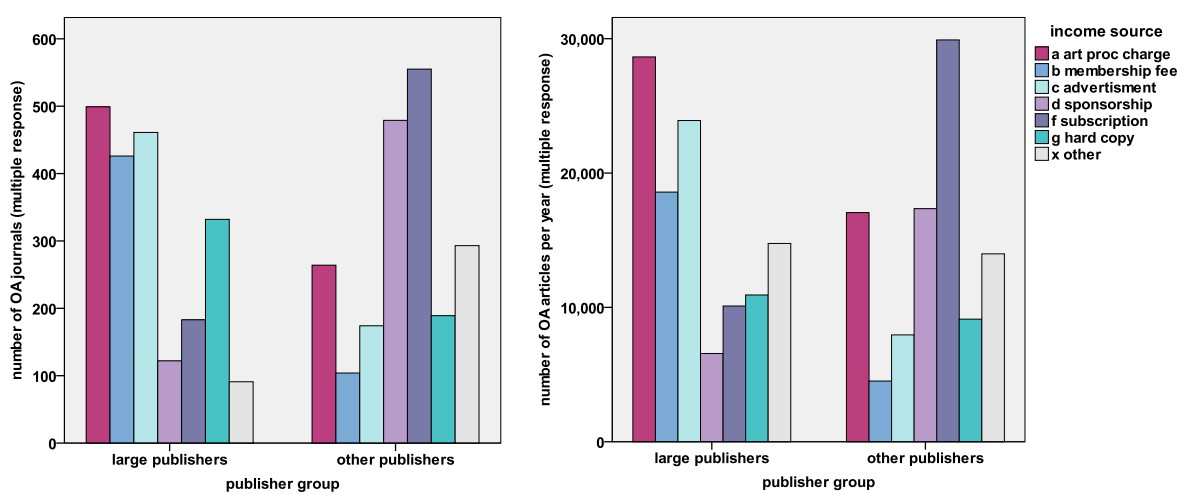

*Figure 2: Number of journals (left) and articles (right) as a function of the income source, for the large publishers and the other publishers.*

The prevalence of article processing charges is correlated with publisher size. About 90% of the publishers with more than 50 journals and 65% of medium sized publishers (10 to 49 journals) mostly rely on article processing charges. This holds only for about 20% of publishers with two to nine journals and 15% with a single journal.

## Copyright and licensing

Half of the large publishers use some version of a Creative Commons license. These seven publishers publish 72% of the titles and 71% of the articles investigated. Out of these articles, 82% are published under the cc-by license and 18% under the more restrictive cc-by-nc license. The other seven large publishers request a transfer of copyright.

Copyright information is available only for 73% of the journals of other publishers, and among these, the transfer of copyright is much more common (69%) than a Creative Commons license (21%). The author retains copyright for the remaining 10% of these journals.

## Impact factor

Inclusion of a journal in the Scopus and ISI-JCR databases are often used as proxy for the journal quality. Out of the journals considered in this study, 41% were included in Scopus and 11% in ISI-JCR. The inclusion in ISI-JCR does not depend either on the size of the publisher or on the volume of articles of the journal. It does depend, however, on the subject area: only 19 journals in SSH, accounting together for less than 500 articles in a year, appear in ISI-JCR. This trend is common also for traditional journals. It should be noted that several journals considered in this study are relatively new and therefore not yet included in ISI-JCR, which has a time-lapse of about three years.

## The uptake of the hybrid model

The twelve large publishers offering the hybrid option are listed in the first column of Table 4, together with their number of journals, with and without a hybrid option. These publishers publish 8,100 journals mostly in STM, and representing one third of all STM journals currently published [9]. A quarter of these journals offer a hybrid option. Table 4 also presents the number of open access articles published in journals with the hybrid option in a given time period. This number is compared to the corresponding total article output and the article output in hybrid journals. Some estimates were necessary as data were not always complete. The total number of articles, when not available, is estimated from commonly used databases. The number of articles in hybrid journals, when not available, is estimated by assuming a constant number of articles per journal and using the fraction of journals offering a hybrid option for that given publisher.

| Publisher | Journals without hybrid option | Journals with hybrid option | Time range | OA articles | Total articles | Articles in hybrid journals only |
|---|---|---|---|---|---|---|
| American Chemical Society | 0 | 35 | Jan - Dec 2009 | 210 | 34,611 | 34,611 |
| American Physical Society | 0 | 7 | Jan - Jun 2009 | 12 | 9,558 | estimate 9,400 |
| Cambridge University Press | 238 | 15 | Jan 2008- Jun 2009 | 22 | estimate 15,000 | estimate 900 |
| Elsevier (incl. Cell Press) | 2,310 | 68 | Jan - Oct 2009 | 430 | estimate 202,000 | estimate 21,250 |
| Nature Publishing Group | 72 | 14 | Jan - Nov 2009 | 147 | estimate 12,000 | 2,693 |
| Oxford University Press | 147 | 90 | 2008 | 882 | 13,241 | estimate 1,200 |
| PNAS | 0 | 1 | Jan - Nov 2009 | 840 | 3,253 | 3,253 |
| Royal Society (UK) | 0 | 7 | Jan - Oct 2009 | 143 | 1,823 | 1,823 |
| SAGE | 560 | 54 | 2009 | 10 | 25,631 | 5,147 |
| Springer | 690 | 1,100 | 2009 | 1,520 | 157,000 | 100,000 |
| Taylor&Francis | 1,000 | 300 | 2008 | 24 | 60,000 | estimate 15,000 |
| Wiley Blackwell | 1,100 | 300 | Jan - Oct 2009 | 342 | estimate 112,000 | estimate 24,000 |

*Table 4: Overview of hybrid offer and uptake for the largest publisher offering this model*

Over a time period of 12 months, the total number of hybrid open access articles divided by the total number of articles results in an average open access article share of about 0.7% across the 12 publishers. This relatively low rate reflects the fact that only a quarter of the journals of these publishers offer a hybrid option. To eliminate this bias, the ratio is calculated again only for journals offering a hybrid option. The measured open access share in hybrid journals is found to be around 2%.

## 4. Discussion

The analyses presented above aim to further understanding of open access publishing opportunities. A similar approach has been followed before. As an example, Kaufman-Wills (2005) [10], Dewatripont (2006) [11], Regazzi (2004) [12], Morris (2006) [13] used data from the DOAJ in their studies addressing open access journals, number of articles for journals indexed in ISI-JCR, frequency of use of an article processing fee. Our results augment the existing body of knowledge:
- article level information was collected for journals indexed not only in ISI-JCR or SCOPUS but for a wider set;
- income sources for sustaining a functional operation were investigated beyond the article-processing-charge, the main focus of previous analyses;
- copyright/licensing practices of journals were analysed;
- correlations between the attribution of impact factors and publisher characteristics, were considered for the first time.

Known limitations of this study are the facts that only journals in English were considered; that manual analysis of web-pages may have introduced small systematic uncertainties; and that only the largest publishers with a hybrid option were contacted to provide information on the number of articles published.

The main findings of this first phase of the SOAP project are summarised as follows:
- There are at least 120,000 open access articles published each year in fully open access journals or hybrid journals.
- Each year of the last decade saw the launch of 200-300 new open access journals, with a peak in 2007-2008 due to the activities of Bentham and Hindawi.
- The distribution of journals per publishers is extremely skewed. A small number of "large publishers" publish a large number of journals and articles. A vast majority of publishers with has a single journal.
- "Large publishers" are predominantly active in the STM subject fields and are more likely to be commercial companies rather than not-for-profit.
- The distribution of open access journals over disciplines is rather even. Grouped together, however, two thirds of the journals and three quarters of the articles are in STM.
- Large publishers are more likely to rely on article processing charges (as well as membership fees and advertising) whereas the smaller publishers base their business more on sponsorship and subscriptions, in addition to article processing charges.
- Both large and smaller publishers are equally likely to have journals with an impact factor.
- Large publishers mostly use a version of Creative Commons licensing while several smaller publishers request the transfer of copyright to the publisher.
- Twelve large publishers with a total of about 8,100 journals offer a hybrid option for 25% of their titles.
- Where a hybrid option is presented, about 2% of the articles are published using this option.

**Acknowledgements**
We are indebted to: the staff at the DOAJ for making their data available for this study; the publishers who responded to our inquiry about open access articles in hybrid journals - the American Physical Society, Cambridge University Press, Elsevier, Nature Publishing Group, Oxford University Press, PNAS, Royal Society (UK), SAGE, Springer, Taylor & Francis and Wiley Blackwell; Margit Palzenberger for conducting the statistical analysis; Wolfgang Kurtz for his comments on our manuscripts; and Bo-Christer Björk for his suggestions. This work would not have been possible without the efforts of Johannes Breimeier, Sylvia Cremer, Jenny Drey, Julia Graepel, Volker Kruppa, Tina Planck and Teodora Todorova in the data collection. The research results of this project are co-funded by the European Commission under the FP7 Research Infrastructures Grant Agreement Nr. 230220.